\documentstyle[twocolumn,aps,epsf,prb]{revtex}

\begin{document} 
\draft 

\twocolumn[
\hsize\textwidth\columnwidth\hsize\csname@twocolumnfalse\endcsname

\title{Continuous weak measurement of the macroscopic quantum 
coherent oscillations } 

\author{D.V. Averin}

\address{Department of Physics and Astronomy, SUNY at Stony Brook, 
Stony Brook, NY 11794-3800} 

\date{\today} 
 
\maketitle 
 
\begin{abstract} 

The problem of continuous quantum measurement of coherent 
oscillations in an individual quantum two-state system is studied 
for a generic 
model of the measuring device. It is shown that for a symmetric 
detector, the signal-to-noise ratio of the measurement, defined 
as the ratio of the amplitude of the oscillation line in the 
output spectrum to background noise, is independent of the 
coupling strength between oscillations and the detector, and is 
equal to $(\hbar/\epsilon)^2$, where $\epsilon$ is the detector 
energy sensitivity. The fundamental quantum limit of 4 imposed by 
this result on the signal-to-noise ratio of the measurement with 
an ``ideal'' quantum-limited detector reflects the general 
tendency of a quantum measurement to localize the system in one 
of the eigenstates of the measured observable. These results are 
applied to specific measurements of the quantum oscillations of 
magnetic flux with a dc SQUID, and oscillations of charge measured 
with a Cooper-pair electrometer. They are also used to calculate 
the energy sensitivity of a quantum point contact as detector.

\end{abstract} 

\vspace*{4ex}

] 

\section{Introduction}   

\vspace{1ex} 
 
Quantum superposition of macroscopically distinct states is one 
of the most characteristic features of quantum behavior at 
the macroscopic level. In the ``mesoscopic'' regime, when the 
states involved in the superposition correspond to a collective 
motion of a number of particles that is larger than one but not 
quite macroscopic, the superposition of states has been 
demonstrated for photons in a high-quality microwave cavity 
\cite{b1}, and for the center-of-mass motion of large molecules 
\cite{b2}. Among the most basic dynamic manifestations of quantum 
superposition of states are quantum coherent oscillations between 
the two basis states of a two-state system. However, while the 
macroscopic quantum phenomena brought about by the incoherent 
quantum tunneling are by now commonly found in a variety of 
systems ranging from mesoscopic tunnel junctions in the regimes 
of flux \cite{b3,b4} and charge \cite{b5,b6} dynamics, to 
molecular \cite{b7,b8} and nano-magnets \cite{b9}, the situation 
with experimental observation of macroscopic quantum coherent 
(MQC) oscillations remains much more uncertain. 
Claim of the observation of the MQC oscillations in molecular 
magnets \cite{b12} remain highly controversial \cite{b13,b14}. 
Remarkable experimental demonstration \cite{b10} of the MQC 
oscillations in the charge-dynamics regime of a small Josephson 
junction is open to criticism that the two charge states of the 
observed quantum superposition differ in charge only by the the 
charge of one Cooper pair, and the oscillations between these 
two states can not be interpreted as macroscopic. Although this 
criticism is not fully justified, since the charge dynamics of 
a small Josephson junction is just another representation of 
its flux dynamics, which is the paradigm of the ``macroscopic'' 
quantum dynamics \cite{b11}, it does not allow to consider the 
question of MQC oscillations to be completely settled. 

Recently, macroscopic quantum dynamics has attracted renewed 
attention as the possible basis for development of scalable 
quantum logic circuits for quantum computation. In this context, 
macroscopic quantum two-state system plays the role of a qubit, 
an elementary building block of a quantum computer. Several 
variants of qubits and quantum logic gates have been proposed 
\cite{b15,b16,b17,b18} that are based on the macroscopic quantum 
dynamics of Josephson junctions. Many characteristics of the 
macroscopic qubits compare favorably with those of the microscopic 
qubits: they are insensitive to disorder at the microscopic level   
and offer much larger freedom in design and fabrication of complex 
systems of qubits. The price of these advantages is the problem of 
the environment-induced decoherence, which is typically much more 
serious for the macroscopic than microscopic quantum systems. 
Suppression of decoherence to the acceptable level requires 
thorough isolation of the qubit from its environment, the condition 
that typically limits the ability to control the qubit dynamics. 

This trade-off between the external control and decoherence in a 
qubit has a fundamental aspect related to measurement. Even in the 
case of a perfect set-up, the measurement necessary to observe the 
MQC oscillations perturbs the system by projecting its state on the 
eigenstates of the measured observable, and  therefore presents 
an unavoidable source of decoherence. The intensity of this 
measurement-induced decoherence increases with increasing 
coupling strength between the detector and the oscillations, and 
more efficient measurement leads to stronger decoherence. First 
approaches to the problem of measurement of the MQC oscillations 
in an individual two-states system \cite{b11,b19,b20,b21} 
suggested for oscillations of magnetic flux in SQUIDs, considered 
only the conventional limit of strong or ``projective'' quantum 
measurements, in which the detector-oscillation coupling is 
strong. In this case, the measurement leads to rapid localization 
of the measured observable (flux) in one of its eigenstates, and 
suppresses the oscillation. This means that the time evolution of 
oscillations can be studied with strong measurements only if the 
detector can be switched on and off on the time scale shorter that 
the oscillations period, and only in the ``ensemble'' of 
measurements, i.e. when the experiment is repeated many times 
with the same initial conditions. The measurement cycle consists 
then of preparation of the initial state of the system followed 
by its free evolution and the subsequent measurement. The 
information about dynamics of the oscillation is contained in the 
probability distribution of the measurement outcomes. Since the 
oscillation frequency is limited from below by several factors 
including the decoherence rate and temperature, the need to 
switch the detector on and off rapidly in this approach presents 
at the very least a serious technical challenge. 

The goal of this work is to study quantitatively a new approach 
to measurement of the MQC oscillations that is based on weak 
quantum measurements \cite{b22,b23,b24}, in which the dynamic 
interaction between the detector and the measured system is weak 
and does not establish perfect correlation between their states. 
Such a weak measurement provides only limited information about 
the system but, in contrast to strong measurements, perturbs the 
system only slightly and can be performed continuously. In this 
work, the process of continuous weak 
measurement of the MQC oscillations in an individual two-state 
system is considered quantitatively. Recent results \cite{b25} 
for continuous measurements of electron oscillations in coupled 
quantum dots by a quantum point-contact are reformulated within 
a generic detector model and applied explicitly to the MQC 
oscillations of flux measured by a dc SQUID and oscillations of 
charge measured with a Cooper-pair electrometer. It is shown 
that the signal-to-noise ratio of the measurement, defined as 
the ratio of the amplitude of the oscillation line in the 
output spectrum of the detector to the 
background noise, is fundamentally limited by the trade-off 
between the acquisition of information and dephasing due to 
detector backaction on the oscillations. This limitation is 
the least restrictive for a symmetric detector, for which the 
signal-to-noise ratio can be expressed as $(\hbar/\epsilon)^2$, 
where $\epsilon$ is the detector energy sensitivity. Since 
the energy sensitivity $\epsilon$ is limited for regular 
(non-QND) quantum measurements by $\hbar/2$, the signal-to-noise 
ratio of the continuous weak measurement of the quantum 
coherent oscillations is limited by 4. This limit reflects the 
fundamental tendency of quantum measurement to localize the 
system in one of the eigenstates of the measured observable. 
As a spin-off, the established relation between the 
signal-to-noise ratio and energy sensitivity is used to 
demonstrate that the quantum point contact is the quantum-limited 
detector with energy sensitivity $\epsilon=\hbar/2$.

\section{Continuos measurement of the MQC oscillations with a 
linear detector}   

\vspace{1ex}

We consider the MQC oscillations in an individual 
two-states system, with the two basis states separated in energy by 
$\varepsilon$ and coupled by the tunneling amplitude $-\Delta/2$. 
The basis states are chosen to coincide with the eigenstates of an 
oscillating variable $x$, for instance, magnetic flux in a SQUID 
loop. In this basis, $x=x_0 \sigma_z/2$, where $x_0$ is the 
difference between the values of $x$ in the two states of the 
system, and $\sigma_z$ is the Pauli matrix. In the simplest  
measurement scheme considered in this work, the detector measures 
directly the oscillating variable $x$, and therefore is coupled 
to $x$. This means that the Hamiltonian describing the measurement 
set-up (Fig.\ 1) is: 
\begin{equation}
H=-\frac{1}{2}(\varepsilon \sigma_z +\Delta \sigma_x +\sigma_z f)  
+ H_0\, ,
\label{1} \end{equation} 
where $H_0$ is the Hamiltonian of the detector, and $f$ is the 
detector operator that couples it to the oscillations. For 
convenience, the amplitude $x_0$ of the oscillations and the 
coupling strength are included in $f$.   

\begin{figure}[htb]
\setlength{\unitlength}{1.0in}
\begin{picture}(3.3,.9) 
\put(-0.05,0.15){\epsfxsize=3.3in\epsfysize=0.7in\epsfbox{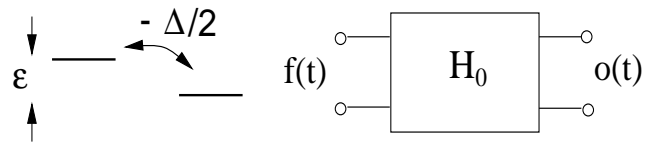}}
\end{picture}
\caption{The diagram of the MQC oscillations in an individual 
two-state system measured by a linear detector. Detector with 
the Hamiltonian $H_0$ is weakly coupled through the operator 
$f$ to the oscillating coordinate of the system. The detector 
output $o(t)$ consists of noise and linear response to the input 
oscillations. }

\end{figure}

We assume that the characteristic response time of the detector 
is short, e.g., much shorter than the period of measured 
oscillations, and that the detector operates in the linear regime. 
These two assumptions mean that the detector output $o(t)$ can 
be written as 
\begin{equation} 
o(t)=q(t) + \frac{\lambda}{2} \sigma_z(t) \, , 
\label{2} \end{equation}  
where $q(t)$ is the noise part of the output, and $\lambda$ is a 
response coefficient. The linearity assumption implies that the 
coupling $-\sigma_z f/2$ of the detector to the oscillations is 
sufficiently weak so that the detector response can be described 
in the linear-response approximation: 
\begin{equation} 
\lambda = i \int_0^{\infty} d\tau e^{i\omega \tau} 
\langle [q(\tau),f] \rangle_0 \, .  
\label{3} \end{equation}  
The average $\langle \ldots \rangle_0$ in eq.\ (\ref{3}) is taken 
over the stationary density matrix of the detector. Since the 
response coefficient $\lambda$ is independent of frequency $\omega$, 
eq.\ (\ref{3}) can be written in terms of the $q$--$f$ correlator 
as 
\begin{equation} 
\langle q(t+\tau)f(t) \rangle_0 = 2\pi S_{qf} \delta (\tau-0)\, , 
\;\;\; S_{qf}\equiv \frac{a-i\lambda}{4\pi} \, , 
\label{4} \end{equation}
where the infinitesimal shift in the argument of the 
$\delta$-function represents small but finite response time of the 
detector, and is needed to resolve the ambiguity in eq.\ (\ref{3}). 
Parameter $a$ in eq.\ (\ref{4}) is introduced to represent the 
real part of the $q$--$f$ correlator that is not determined by 
the response coefficient $\lambda$. 

The role of detector in a quantum measurement is to convert 
the quantum input signal, in our case, the oscillations $x(t)=x_0 
\sigma_z(t)/2$, into the output signal that is already classical 
and can be dealt with (e.g., monitored or recordered) without 
``fundamental'' problems. Condition of the classical behavior of 
the detector output requires the output spectral density to be 
much larger than the spectral density of the zero-point 
fluctuations in the relevant range of low frequencies of the 
input signal. For a detector with a short internal time scale 
this means that the noise $q(t)$ is $\delta$-correlated on the 
time scale of the input signal: 
\begin{equation} 
\langle q(t+\tau)q(t) \rangle_0 = 2\pi S_q \delta(\tau)\, .  
\label{5} \end{equation}
Here $S_q$ is the constant low-frequency part of the spectral 
density $S_q(\omega)$ of the detector output noise. 

For a quantum-limited detector at small temperature $T\rightarrow 
0$, eqs.\ (\ref{4}) and (\ref{5}) impose a constraint on the 
spectral density $S_f$ of the coupling operator $f$. Indeed, 
eqs.\ (\ref{4}) and (\ref{5}) can be written explicitly in the 
basis of the energy eigenstates $|k\rangle$ of the detector and 
give the following expression for the spectral density $S_q$: 
\begin{equation} 
S_q=\int d \varepsilon_k d \varepsilon_{k'} \nu (\varepsilon_k)  
\nu (\varepsilon_{k'}) \rho_k \langle k| q|k'\rangle \langle k'|q| 
k\rangle \delta(\varepsilon_k - \varepsilon_k' -\omega) \, . 
\label{6} \end{equation}
Expression for the correlation amplitude $S_{qf}$ is similar, with 
the matrix element $\langle k'|q|k \rangle$ replaced by the matrix 
element of $f$. In these expressions, $\varepsilon_k$ is the energy 
of the state $|k\rangle$, $\rho_k$ is the probability to be in this 
state, and $\nu$ is the state energy density. Since $S_q$ is 
independent of the frequency $\omega$, eq.\ (\ref{6}) is satisfied 
when the matrix elements of $q$ and the density of states are 
constant, and eq.\ (\ref{6}) can then be written as 
\[ S_q= |\langle q\rangle |^2 \nu^2 \, .  \] 
It should be noted that the constant matrix elements $\langle q 
\rangle$ and $\langle f\rangle$ are off-diagonal in the $k$-basis 
and can be imaginary. Following the same steps for $S_f$ we 
express this spectral density in terms of $S_{q}$ and $S_{qf}$: 
\begin{equation} 
S_f= |\langle f\rangle |^2 \nu^2 = |S_{qf}|^2/S_q  \, . 
\label{8} \end{equation}
Equations (\ref{8}) and (\ref{4}) relate the backaction noise of 
the detector determined by the spectral density $S_f$ to its 
response coefficient and the output noise. Conceptually, such a 
relation resembles the fluctuation-dissipation theorem that links  
response of the system to its equilibrium fluctuations, but it 
does not have the status of a ``theorem''.  It is obvious from the 
derivation above that eq.\ (\ref{8}) is not necessarily valid for 
an arbitrary system playing the role of detector in a quantum 
measurement. Nevertheless, it holds for several of the ``standard'' 
detectors: quantum point contact, resistively-shunted dc SQUID, 
and Cooper-pair electrometer, considered later in this work.  

Making use of $q$-- and $f$-- correlators, we can calculate the 
spectral density of the detector output $o(t)$ in the process of 
continuos measurement of the MQC oscillations. From eq.\ (\ref{2}), 
the correlation function of $o(t)$ is: 
\begin{equation} 
K_o(\tau)= 2\pi S_q \delta (\tau) + \frac{\lambda^2}{4} \mbox{Tr} 
\{ \rho \sigma_z \sigma_z(\tau) \} \, ,  
\label{10} \end{equation} 
where $\rho$ is the stationary density matrix of the 
two-state system established as a result of the interaction with 
the detector. Averaging the Heisenberg equation of motion of the 
operator $\sigma_z(\tau)$ over the $\delta$-correlated backaction 
noise $f$ of the detector, we get the set of equations for the time 
evolution of the matrix elements $\sigma_{ij}$ of $\sigma_z (\tau)$: 
\begin{equation}
\dot{\sigma}_{11}=  \Delta \,\mbox{Im} \sigma_{12}\, , \;\;\;\; 
{\dot\sigma}_{12}=  (i\varepsilon -\Gamma ) \sigma_{12} - i\Delta  
\, \sigma_{11} \, ,  
\label{9} \end{equation}
and $\sigma_{22}=-\sigma_{11}$, with the rate 
\begin{equation} \Gamma=\pi S_f \label{12} \end{equation}   
describing the backaction dephasing of the oscillations by the 
detector. The density matrix $\rho$ of the two-state system 
satisfies the same set of equations (\ref{9}), except for the 
normalization, $\rho_{11}+\rho_{22}=1$, and its stationary value 
is $\rho=1/2$. Solving eqs.\ (\ref{9}) with the initial condition 
$\sigma_z (0)= \sigma_z$ and averaging $\sigma_z\sigma_z (\tau)$ 
over $\rho=1/2$ we find the spectral density $S_o (\omega ) = 
(1/2\pi) \int_{-\infty}^\infty K_o(\tau) e^{i \omega \tau } d\tau$.  
Under the conditions of ``resonance'', $\varepsilon=0$, when the 
oscillation amplitude is maximum, we get: 
\begin{equation}
S_o(\omega ) = S_q +\frac{\Gamma \lambda^2}{4\pi} 
\frac{ \Delta^2 } {(\omega^2-\Delta^2)^2+\Gamma^2\omega^2} \, . 
\label{11} \end{equation}

When $\varepsilon \neq 0$, it is convenient to calculate the spectrum 
numerically from eq.\ (\ref{9}). The spectrum in this case is plotted 
in Fig.\ 2 for several values of $\varepsilon$ and the dephasing rate 
$\Gamma$. For weak dephasing, $\Gamma \ll \Delta$, the spectrum 
consists of a zero-frequency Lorentzian that vanishes at 
$\varepsilon=0$ and grows with increasing $|\varepsilon |$, and a 
peak at the oscillation frequency $\Omega= (\Delta^2+ \varepsilon^2 
)^{1/2}$. The peak at zero frequency reflects the incoherent 
transitions with a small rate of order $\Gamma$ between the states 
of the two-state system. The high-frequency peak of the MQC 
oscillations also has the width $\Gamma$. While this width can be 
small for sufficiently weak dot-contact coupling, the height of the 
oscillation peak cannot be arbitrarily large in comparison to the 
background noise spectral density $S_q$. At $\varepsilon=0$, 
when the amplitude of the oscillations is maximum, the peak height 
is $S_{max}=\lambda^2/4\pi \Gamma$. Even in this case, the ratio of 
the peak height to the background is limited: 
\begin{equation} 
\frac{S_{max}}{S_q} = \frac{\lambda^2}{4\pi^2S_fS_q} = 
\frac{4 \lambda^2}{\lambda^2+a^2} \leq 4 \, . 
\label{15} \end{equation}

This limitation is universal, e.g., independent of the coupling 
strength between the detector and oscillations, and reflects 
quantitatively the interplay between measurement of the MQC 
oscillations and their backaction dephasing. The fact that the 
height of the spectral line of the oscillations can not be much 
larger than the noise background means that, in the time domain, 
the oscillations are drowned in the shot noise. 

\begin{figure}[htb]
\setlength{\unitlength}{1.0in}
\begin{picture}(3.,2.4) 
\put(.0,.0){\epsfxsize=3.in\epsfysize=2.4in\epsfbox{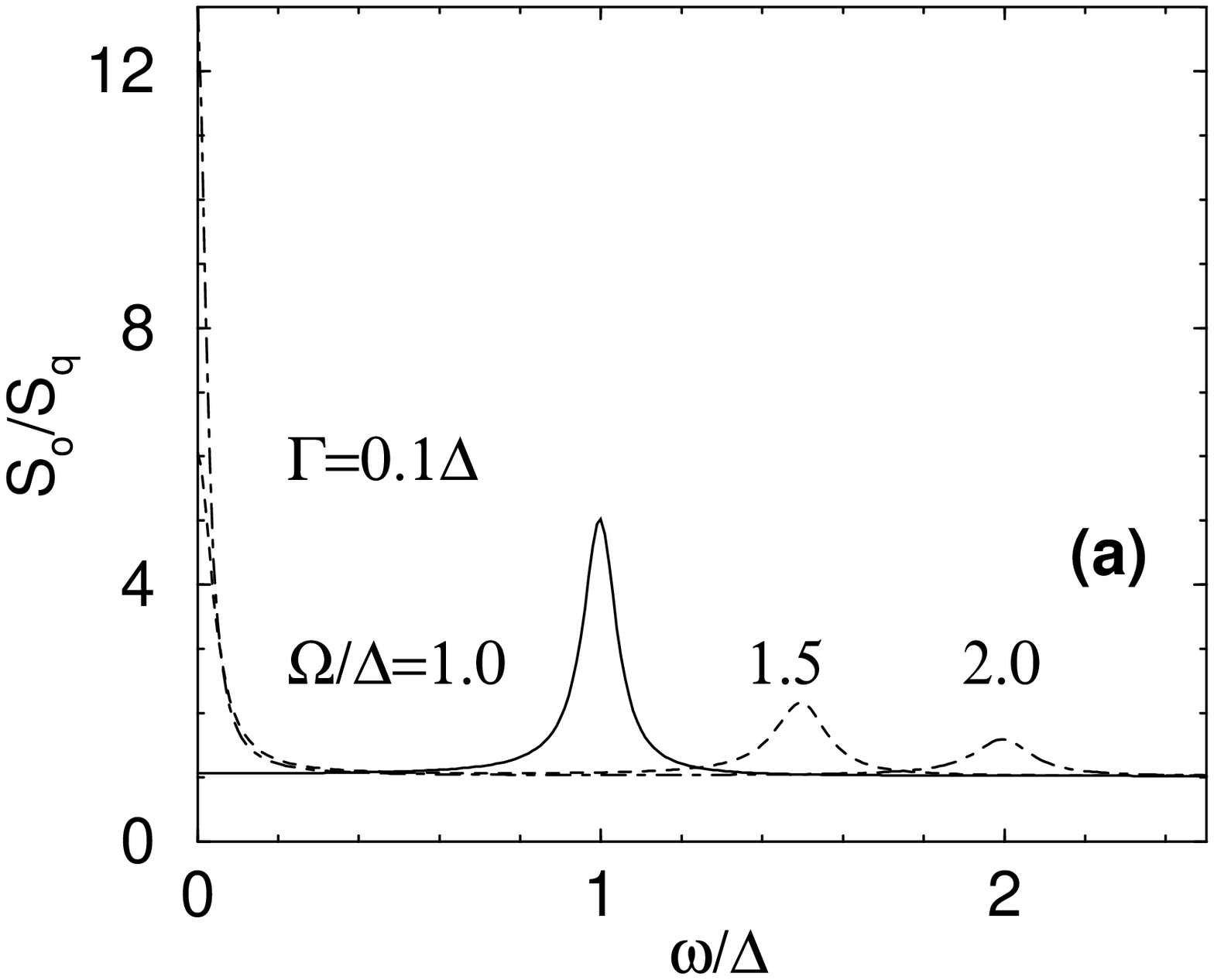}}
\end{picture}

\begin{picture}(3.0,2.4) 
\put(.0,.0){\epsfxsize=3.0in\epsfysize=2.4in\epsfbox{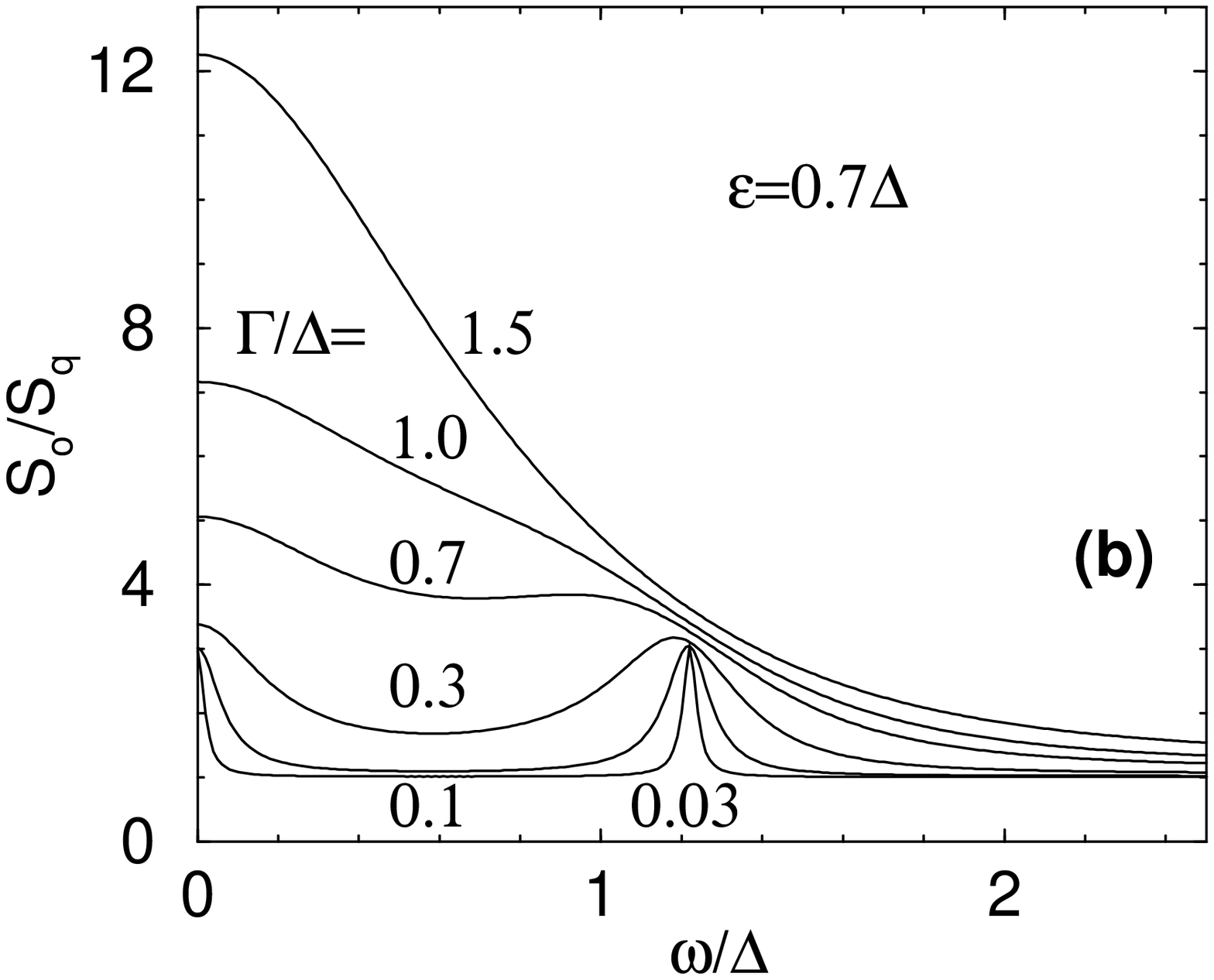}}
\end{picture}

\caption{Spectral density $S_o(\omega )$ of the output of the 
symmetric detector with minimal backaction dephasing ($a=0$) 
measuring the MQC oscillations for several values of (a) the 
energy bias $\varepsilon$ reflected in the oscillation frequency 
$\Omega=(\Delta^2+ \varepsilon^2)^{1/2}$, and (b) the rate $\Gamma$ 
\protect (\ref{12}) of the measurement-induced dephasing. }

\end{figure} 
When the backaction dephasing rate $\Gamma$ increases, the 
oscillation line broadens towards the lower frequencies, and 
eventually turns into the growing spectral peak at zero frequency 
associated with the incoherent jumps between the two basis states 
of the two-state system. At large $\Gamma$, when the coherent 
oscillations are suppressed, the rate of incoherent tunneling 
decreases with increasing $\Gamma$. For instance, at $\Gamma \gg 
\Omega$, the tunneling rate is $\gamma=\Delta^2/ 2\Gamma$ , and 
the spectral density of the detector response at low frequencies 
$\omega \! \sim \! \gamma$ has the standard Lorentzian form, 
$S_o(\omega ) -S_q=2 \gamma \lambda^2/4\pi (4\gamma^2+ \omega^2)$. 
Suppression of the tunneling rate $\gamma$ 
with increasing dephasing rate $\Gamma$ is an example of the generic 
``Quantum Zeno Effect'' in which quantum measurement suppresses the 
decay rate of a metastable state. In the context of search for the 
macroscopic quantum coherent oscillations, the Lorentzian spectral 
density has been observed and used for measuring the tunneling rate 
of incoherent quantum flux tunneling in SQUIDs \cite{b26}. 
 
The maximum signal-to-noise ratio $S_{max}/S_q$ (\ref{15}) 
is attained if the fundamental backaction of the detector is the 
only mechanism of dephasing of the coherent oscillations. We now 
discuss briefly the effect of a weak additional dephasing and 
energy-relaxation on the spectral density of the 
oscillations. The efect of such a weak relaxation is noticeable if 
the backaction dephasing is also weak, $\Gamma \ll \Delta$. Energy 
relaxation arises typically due to interaction with some external 
system (``reservoir'') that is in equilibrium at 
temperature $T$. The interaction term in the Hamiltonian 
can be written similarly to the interaction with the detector 
(\ref{1}) as 
\begin{equation} 
H_c = -\sigma_z f_r \, , 
\label{16} \end{equation}
where $f_r$ is the reservoir force acting on the system. Under 
the assumption of the frequency-independent relaxation rate, the 
standard free equilibrium correlator of this force is (see, e.g., 
\cite{b27}): 
\begin{equation}  
\langle f_r(t) f_r(t+\tau) \rangle = \alpha \int \frac{d 
\omega }{\pi} \frac{ \omega e^{i\omega \tau } }{1-e^{- \omega /T} 
} \, , \label{cor} \end{equation} 
where parameter $\alpha$ characterizes the relaxation strength. 
Comparison of this correlator with the $\delta$-correlated 
backaction noise of the detector shows that the detector (\ref{1}) 
is acting effectively as a reservoir with temperature that is much 
larger that the energies of the two-state system.   

Energy relaxation makes the stationary average value of $\sigma_z$ 
non-vanishing, and the output correlation function should now be 
calculated as 
\begin{equation} 
K_o(\tau)= K_q (\tau) + \frac{\lambda^2}{8} [\langle \sigma_z 
\sigma_z(\tau) +\sigma_z(\tau) \sigma_z\rangle - 2 \langle 
\sigma_z \rangle^2 ] \, .  
\label{cor2} \end{equation}
For weak coupling, it is convenient to find the time evolution of 
$\sigma_z(\tau)$ in the basis of eigenstates of the two-state 
Hamiltonian. In this basis, the Hamiltonian including interaction 
with the reservoir (and omitting temporarily the detector) 
is : 
\begin{equation} 
H = -\frac{1}{2} \Omega \sigma_z - \frac{1}{\Omega }(\varepsilon 
\sigma_z - \Delta \sigma_x)f_r +H_r\, . 
\label{17} \end{equation}
$H_r$ is the Hamiltonian of the reservoir. 
Heisenberg equations of motion that follow from the Hamiltonian 
(\ref{17}) are: 
\[ \dot{f}_r=i[H_r,f_r] \, , \;\;\;\; \dot{H}_r= \frac{i}{\Omega }
(\Delta \sigma_x-\varepsilon \sigma_z )[f_r,H_r] \, , \]
\[ \dot{\sigma}_z=2f_r\Delta \sigma_y \, , \;\;\;\;
\dot{\sigma}_\pm= \mp i (\Omega +2\varepsilon f_r/\Omega)\sigma_\pm
\mp i \Delta \sigma_z/\Omega \, , \]
where $\sigma_\pm\equiv (\sigma_x\pm i\sigma_y)/2$. Integration of 
the first two of these equations gives in the first order in coupling 
(\ref{16}) to the two-state system: 
\[ f_r(t)=f_r^{(0)}(t)-i\Delta \int^t \! d\tau \int^\tau \! d\tau' 
[f_r^{(0)}(\tau),f_r^{(0)}(\tau')] \sigma_y(\tau') \, , \] 
where $f_r^{(0)}$ is the free part of the fluctuating reservoir 
force in absence of coupling. Solving the second pair of the 
Heisenberg equations up to the second order in coupling, making the 
rotating-wave approximation, and tracing out 
the reservoir degrees of freedom with the help of the correlator 
(\ref{cor}), we get a set of equations for the evolution of the 
matrix elements $s_{ij}$ of the operator of the oscillating variable 
(given by $\sigma_z(\tau)$ in the original ``position'' basis) in the 
eigenstate basis: 
\[ \dot{s}_{jj}(\tau) = \Gamma_e [\frac{\varepsilon}{\Omega} -\coth \{ 
\frac{\Omega}{2T} \} s_{jj}] + (-1)^j \frac{\Gamma \Delta^2}{2\Omega^2} 
(s_{11}-s_{22}) \, , \]

\vspace*{-1ex} 

\begin{equation}
{\dot s}_{12}(\tau) =(i\varepsilon -\Gamma_0 ) s_{12} \, . 
\label{21} \end{equation}
Initial conditions for these equations are: 
$s_{11}=-s_{22}=\varepsilon/\Omega$, and $s_{12}=-\Delta/\Omega$. 
The characteristic energy-relaxation rate 
in eq.\ (\ref{21}) is $\Gamma_e =2\alpha \Delta^2/\Omega$, and the 
total dephasing rate is 
\[ \Gamma_0= \frac{1}{\Omega^2} [\alpha (\Delta^2 \Omega \coth \{ 
\frac{\Omega}{2T} \} +4\varepsilon^2 T ) + \Gamma (\varepsilon^2+ 
\frac{\Delta^2}{2})] \, .\] 
In eqs.\ (\ref{21}), we also added the detector dephasing terms 
from eq.\ (\ref{9}) ``rotated'' from the position basis into the 
eigenstate basis. 

The density matrix $r$ of the two-state system in the basis of 
eigenstates satisfies similar equations, and the stationary values 
of its matrix elements are $r_{12}=0$ and $r_{11}=(\Gamma_t+ 
\Gamma_e)/2 \Gamma_t$, where 
\[ \Gamma_t\equiv \Gamma_e \coth (\Omega/2T)+\Gamma 
\Delta^2/\Omega^2 \, . \] 
Using these relations, the definition (\ref{cor2}), and the 
evolution equations (\ref{21}) we find the spectral density: 
\begin{eqnarray}
\lefteqn{
S_o(\omega )=S_q + \frac{\lambda^2 }{4\pi \Omega^2} \times  }
\nonumber \\  
& & \left( [1-(\frac{\Gamma_e}{\Gamma_t})^2] \frac{2 \varepsilon^2 
\Gamma_t }{\omega^2+ \Gamma_t^2} + \sum_{\pm}  \frac{\Delta^2 
\Gamma_0} {(\omega \pm\Omega )^2+\Gamma_0^2 } \right) \, . 
\label{23} \end{eqnarray}
As in the case without energy relaxation, the spectral density 
consists of a zero-frequency Lorentzian of width $\Gamma_t$ and 
peaks at $\pm \Omega$ of width $\Gamma_0$ due to coherent 
oscillations. For weak relaxation, the incoherent slow 
transitions giving rise to the low-frequency noise are the 
transitions between the two energy eigenstates of the system. 
The height of the oscillation peak is suppressed in presence of 
additional energy relaxation that contributes to the dephasing 
rate $\Gamma_e$, and the relative magnitude of the peak, 
$S_{max}/S_q$ is smaller than its value without the relaxation.

\section{Relation to energy sensitivity}   
 
\vspace{1ex}

The detector characteristics for measurement of the quantum 
coherent oscillations in a two-state system considered in the 
previous Section are related to another detector characteristic 
that is used for measurements of the harmonic signals -- see, 
e.g., examples in \cite{b28,b29,b30,b31}, and sometimes is 
loosely referred to as ``energy sensitivity''. 
It is defined by considering the detector measuring 
a harmonic oscillator with a frequency $\omega_0$ and a small 
relaxation rate $\gamma \ll \omega_0$. In this Section, we 
establish the quantitative relation between the signal-to-noise 
ratio $S_{max}/S_q$ for measurements of the two-state systems  
discussed above with the energy sensitivity used in the 
literature for measurements of harmonic signals.  

The Hamiltonian of the damped harmonic oscillator attached 
to a detector is obtained by replacing the two-state part of 
eqs.\ (\ref{1}) and (\ref{16}) with the corresponding  
oscillator terms: 
\begin{equation} 
H = \frac{M}{2}(\dot{x}^2+\omega_0^2 x^2) -x(f+ f_r)+H_0+H_r \, ,  
\label{ho} \end{equation}
where $M$ is the mass of the oscillator and $x$ is the oscillating 
coordinate. Due to linearity of the system, the Heisenberg equation 
of motion for $x$ that follow from the Hamiltonian (\ref{ho}) 
(see, e.g., \cite{b27}) coincides with the classical equation of 
motion of the damped oscillator: 
\begin{equation} 
\ddot{x}-\gamma \dot{x} +\omega_0^2 x = \frac{1}{M}(f_r(t) + 
f(t))  
\, . 
\label{31} \end{equation} 
As in the previous Section, the random forces $f_r(t)$ and $f(t)$ 
are produced, respectively, by the reservoir responsible for the 
energy relaxation of the oscillator and by the detector. (Although 
the operators $f_r$ and $f$ in eqs.\ (\ref{ho}) and (\ref{31}), as 
well as the transfer coefficient $\lambda$ in eq.\ (\ref{32}) below, 
differ from the corresponding quantities used in Section 2 by a 
normalization factor, this distinction is not made explicit. This 
should not lead to any confusion, since the normalization factor 
drops out of all expression which are compared between the two 
Sections.) In general, the right-hand-side of eq.\ (\ref{31}) 
should also contain an external perturbation that creates a 
``signal'' component of the oscillations $x(t)$. However, for the 
discussion of the detector sensitivity, it is appropriate to treat 
the equilibrium fluctuations of the oscillator driven by the 
reservoir noise $f_r(t)$ (e.g., the zero-point oscillations at 
vanishing temperature $T$) as part of the signal. This allows us 
not to include additional signal terms in eq.\ (\ref{31}). 

The sensitivity of the detector is characterized by the detector 
noise contribution to the spectral density of the output 
$S_o$ reduced to the detector input. Similarly to eq.\ (\ref{2}) the 
output of the detector measuring harmonic oscillator is: 
\begin{equation} 
o(t)=q(t) + \lambda x(t) \, . 
\label{32} \end{equation}  
This equation implies that the detector contribution to the output 
noise comes from the two sources: direct output noise $q(t)$, and 
effect of the backaction noise $f(t)$ on the oscillator coordinate 
$x(t)$. Introducing the dynamic response function $G(t)$ of the 
oscillator: 
\begin{equation} 
x(t) = \int_0^{+\infty} d\tau G(\tau) (f_r(t-\tau)+f(t-\tau))\, ,  
\label{33} \end{equation}  
with $G(\tau) = 0$ for $\tau<0$, we see from eq.\ (\ref{31}) 
that 
\begin{equation} 
G(\omega )= \int d\tau e^{-i\omega \tau} G(\tau)=\frac{1}{M}  
\frac{1}{\omega_0^2-\omega^2+i\gamma\omega} \, .
\label{34} \end{equation}
In terms of the response function, $x(\omega)=G(\omega ) 
(f_r(\omega)+f(\omega))$. 

Since the detector noise $f(t)$ is uncorrelated with the reservoir 
force $f_r(t)$, or in general, any other signal component of $x(t)$, 
we see from eqs.\ (\ref{32}), (\ref{33}), and (\ref{34}) that the 
spectral density of the detector output consists of the two additive 
components: signal, i.e., equilibrium spectral density of the 
harmonic oscillations transformed to the output, and the detector 
noise $S_N (\omega)$: 
\[ S_o(\omega)=\lambda^2 |G(\omega )|^2 S_r (\omega) + 
S_N (\omega) \, , \] 
\begin{equation} 
S_N(\omega )= S_q+\lambda^2 |G(\omega )|^2 S_f+2 \lambda \mbox{Re} 
G(\omega ) \bar{S}_{qf} \, . 
\label{35} \end{equation}
Here $\bar{S}_{qf}$ is the symmetrized correlator of the detector 
output and input noises, $\bar{S}_{qf}=(S_{qf}+S_{fq})/2=a/4\pi$ 
(with $a$ defined in eq.\ (\ref{4})), and $S_r (\omega)$ is the 
spectral density of the reservoir force $f_r$: 
\begin{equation} 
S_r (\omega)=(\gamma \omega M/ 2\pi) \coth (\hbar \omega/2T).  
\label{36} \end{equation}

The noise properties of the detector are better characterized if 
its contribution to the output noise is reduced to the input, i.e. 
instead of $S_N(\omega )$ (\ref{35}) we consider the quantity 
$F(\omega) \equiv S_N(\omega )/\lambda^2 |G(\omega )|^2$. For weak 
damping, $\gamma \ll \omega_0$,  when 
\[ |G(\omega )|^2 \simeq \frac{1}{4\omega_0^2 M^2 
((\omega-\omega_0)^2+ \gamma^2/4)} \, ,\] 
we get
\[ F(\omega)= S_f+S_q (2\omega_0 M/\lambda)^2 
((\omega-\omega_0)^2+ \gamma^2/4)- \] 
\[ \bar{S}_{qf} (4\omega_0M/\lambda) (\omega-\omega_0) \, . \] 
The three terms in this equation scale differently with the strength 
of the detector-oscillator coupling, since $\lambda$ and $\bar{S}_{qf}$ 
are proportional to the first power, while $S_f$ is proportional to 
the second power of the coupling strength. $F(\omega)$ can be 
minimized with respect to the coupling strength and also with respect 
to the small detuning $\omega-\omega_0$ between the signal frequency 
and the oscillator frequency. The minimum is reached when  
\[ \omega-\omega_0 = \frac{\lambda\bar{S}_{qf}}{2\omega_0 M S_q} \, , 
\]
and 
\[ \lambda =  \frac{\gamma \omega_0 M S_q}{S_0} \, , \;\;\;\;  
S_0\equiv (S_q S_f -\bar{S}_{qf}^2)^{1/2} \, , \] 
and is equal to 
\begin{equation} 
F_{min} = 2\gamma \omega_0 M\frac{S_0}{\lambda } \, . 
\label{37} \end{equation}

It is convenient to normalize the reduced noise $F$ in such a way 
that it can be directly compared to the equilibrium fluctuations in 
the oscillator driven by $S_r$. Since at this stage we can already 
neglect small difference between the oscillator frequency and signal 
frequency, both the minimum noise $F_{min}$ (\ref{37}) and the 
equilibrium spectral density $S_r$ (\ref{36}) are proportional to the 
oscillator parameters $\gamma \omega_0 M$. This means that with 
appropriate normalization we can define $F_{min}$ directly in terms of 
the number of quanta added to the signal at frequency $\omega$ by the 
detector. This is achieved by introducing the energy sensitivity as 
\begin{equation} 
\epsilon \equiv \frac{\pi}{\gamma \omega_0 M} F_{min}= 
\frac{2\pi }{\lambda } (S_q S_f -\bar{S}_{qf}^2)^{1/2} \, . 
\label{38} \end{equation}
Equations (\ref{8}) and (\ref{4}) of the previous Section show that 
for a quantum-limited detector
\begin{equation} 
\epsilon =\hbar/2 \, .
\label{39} \end{equation} 
(Note that in this Chapter, $\hbar$ is shown only in some of the 
final results.)  
Equation (\ref{39}) agrees with the conclusion of a general theory 
of quantum linear amplifiers, according to which a phase-insensitive 
linear amplifier adds at least half-a-quantum of noise to the 
amplified signal -- see, e.g., \cite{b32}. This result is related to 
eq.\ (ref{39}) since detector in the measurement process plays the 
role of an amplifier transforming weak quantum input signal into the 
classical output.  

Comparing eq.\ (\ref{38}) for the energy sensitivity with eq.\ 
(\ref{15}) for the signal-to-noise ratio of the measurement of the  
quantum coherent oscillations we see finally that these two quantities 
are closely related. When the output and input noises of the detector 
are uncorrelated, $\bar{S}_{qf}=0$, the relation is simple: 
\begin{equation}  
\frac{S_{max}}{S_q}= (\hbar/\epsilon)^2 \, .
\label{40} \end{equation} 
As will be clear from the examples of specific detector considered 
below, the situation with $\bar{S}_{qf}=0$ can be reasonably referred 
to as ``symmetric detector''. As follows from eq.\ (ref{40}), the 
largest signal-to-noise ratio of 4 is obtained for such a symmetric 
detector in the quantum-limited regime with $\varepsilon =\hbar/2$. 

When the input-output correlation is non-vanishing, the 
signal-to-noise ratio is smaller that the value given by eq.\ 
(\ref{40}), while the energy sensitivity $\epsilon$ for measurement 
of the harmonic signal can still be made equal to $\hbar/2$ by 
optimizing the detuning between the signal and the oscillator. 
Another difference between the measurement of harmonic oscillator 
defining $\epsilon$ and measurement of the two-state system, is that 
the minimum noise (\ref{37}) in the oscillator measurement is 
reached only for optimum detector-oscillator coupling, while the 
maximum signal-to-noise ratio (\ref{15}) for the quantum oscillation 
measurement is independent of the coupling strength to the detector 
as long as the coupling is weak.

\section{Energy sensitivity of a quantum point contact}   

\vspace{1ex}

One of the applications of the results obtained in the previous 
Section is the demonstration that a quantum point contact 
that is frequently used as the detector of electric charge or 
voltage \cite{b33,b33b,b34,b35} can reach the quantum-limited 
regime with ultimate energy sensitivity (\ref{39}). 
The mechanism of operation of a quantum point contact as a 
detector utilizes modifications of the electron transmission 
properties of the contact by the measured voltage \cite{b33}. 
When the contact is biased with a large voltage $V$, changes in 
the electron transmission probability lead to changes in the 
current $I$ flowing through the contact which serve as the 
measurement output. Fluctuations of the electric potential 
in the contact region due to the current flow produce the 
backaction dephasing of the measured object by the point contact 
\cite{b34,b35}. This dephasing was calculated for symmetric contacts 
within different approaches in \cite{b36,b37,b38,b39,b40}. It is 
known \cite{b41,b25} that in the case of measurement of a two-state 
system, when the coupling to the system is symmetric, quantum point 
contact is an ideal detector of the quantum coherent oscillations. 
Such a detector causes the minimum dephasing of the oscillations 
that is consistent with the information acquisition by the 
measurement. For asymmetric coupling, the dephasing by the point 
contact is larger that the fundamental minimum \cite{b25,b35}. 

To calculate the energy sensitivity of the point contact detector, 
we start with the standard Hamiltonian of a single-mode point 
contact. Including a weak additional scattering potential $U(x)$ 
for the point contact electrons which is the input signal of the 
measurement we can write the Hamiltonian as  
\begin{equation}
H= \sum_{ik} \varepsilon_k a^{\dagger}_{ik}a_{ik} + U\, , 
\;\;\; U=\sum_{ij}U_{ij} \sum_{kp} a^{\dagger}_{ik}a_{jp}\, . 
\label{41} \end{equation} 
The operators $a_{ik}$ in this Hamiltonian represent point-contact 
electrons in the two scattering states $i=1,2$ (incident from the 
two contact electrodes) with momentum $k$, and $U_{ij}= \int dx \, 
\psi_i^*(x) U(x) \psi_j(x)$ are the matrix elements of the 
potential $U(x)$ in the basis of the scattering states. Here 
$\psi_i(x)$ is the wavefunction of the scattering state, and $x$ 
is the coordinate along the point contact. Several assumptions are 
made about the contact. The bias energy $eV$ is assumed to be much 
larger than temperature $T$, but much smaller than both the Fermi 
energy in the point contact and the inverse traversal time of the 
contact. This allows us to linearize the 
energy spectrum of the point-contact electrons: $\varepsilon_k= 
v_F k$, where $v_F$ is the Fermi velocity, and neglect the momentum 
dependence of the matrix elements $U_{ij}$. The potential $U(x)$ is 
also assumed to be sufficiently weak and can be treated as perturbation. 
In this regime, the point contact operates as a linear detector, 
and the current response to the perturbation $U$ can be calculated 
in the linear-response approximation. The last assumption is that 
the frequencies of the input signal are much smaller than $eV$, the 
fact that allows to treat $U$ as the static perturbation. 

At frequencies much lower that both $eV$ and inverse traversal 
time of the contact, the current is constant throughout the contact 
and the contact response can be calculated at any point $x$. We 
choose the origin of the coordinate $x$ in such a way that the 
unperturbed scattering potential is effectively symmetric, i.e., 
the reflection amplitudes for both scattering states are the same, 
and then take $x$ to lie in the asymptotic region of the scattering 
states. In this case, the standard expression for the current in 
terms of the electron operators $\Psi(x)$: 
\[ I = \frac{-ie\hbar}{2m} (\Psi^{\dagger} \frac{\partial \Psi}
{\partial x} - \frac{\partial \Psi^{\dagger} }{\partial x}  
\Psi)\, , \;\;\;  \Psi(x)= \sum_{ik} \psi_{ik}(x)a_{ik} \, , \] 
gives for the current operator at $x$: 
\begin{eqnarray} 
I= \frac{e v_F}{L} \sum_{kp} [D(a^{\dagger}_{1k}a_{1p} - 
a^{\dagger}_{2k}a_{2p}) + \nonumber \\ 
i(DR)^{1/2}e^{-i(k-p)|x|} (a^{\dagger}_{1k} 
a_{2p} -a^{\dagger}_{2k} a_{1p}) ] \, .  
\label{42} \end{eqnarray} 
Here $D$ and $R=1-D$ are the transmission and reflection 
probabilities of the point contact, $L$ is a normalization length, 
and the variation of the momentum near the Fermi points (i.e., the 
difference between $k$ and $p$) was neglected everywhere besides the 
phase factor in the second term. The reason for keeping this factor 
will become clear later. 

In the linear-response regime, the current response of the point 
contact is driven by the part of the perturbation $U$ causing 
transitions between the two scattering states $\psi_{1,2}$. As shown 
in the Appendix, the real part of the transition matrix element 
$U_{12}$ is related to the change $\delta D$ of the transmission 
probability of the contact: 
\begin{equation} 
U_{12}= \frac{v_F}{L} \frac{\delta D+iu}{2(DR)^{1/2}} \, , 
\;\;\;\; U_{21} = U_{12}^*\, . 
\label{43} \end{equation} 
The imaginary part of $U_{12}$, expressed through a dimensionless 
parameter $u$ in eq.\ (\ref{4}), does not affect the current $I$. 
Qualitatively, it characterizes the degree of asymmetry in the 
coupling of the quantum dots to the point contact; $u=0$ if the 
perturbation potential $U(x)$ is applied symmetrically with respect 
to the main scattering potential of the point contact. 

In the measurement process, the perturbation $U$ represents the 
coupling operator between the contact and the measured system, 
whereas the current $I$ is the measurement output. As follows from 
the discussion in the previous Section, the energy sensitivity of 
the contact is determined by the correlators of $U$ and $I$, and 
by the response coefficient $\lambda$. Using eqs.\ (\ref{41}), 
(\ref{42}), and (\ref{43}) we can evaluate the correlators 
directly. In the limit of large voltages, $eV\gg T$, when the 
contact noise properties are dominated by the shot noise, we get: 
\begin{eqnarray} 
\langle U(t)U(t+\tau) \rangle_0 =\frac{eV}{4\pi} \,  
\frac{(\delta D)^2+u^2}{DR} \, \delta(\tau) \, , \nonumber \\  
\langle I(t+\tau)I(t) \rangle_0 =\frac{e^3VDR}{\pi}\, 
\delta(\tau) \, , \label{45} \\
\langle U(t)I(t+\tau) \rangle_0 = \frac{e^2V}{2\pi} 
(i\delta D+u) \, \delta(\tau-\eta)\, . \nonumber
\end{eqnarray} 
The time delay $\eta \equiv |x|/v_F$ in the last of eqs.\ (\ref{45}) 
comes from the phase factor $e^{-i(k-p)|x|}$ kept in eq.\ (\ref{42}), 
and is infinitesimally small for small traversal time of the contact. 
It is nevertheless important for correct calculation of the contact 
response $\lambda$. From the $U$--$I$ correlator and the standard 
expression for the linear response (\ref{3}) we confirm that 
$\lambda$ is equal to the change of current through the contact due 
to change $\delta D$ of the transmission coefficient, $\lambda=e^2V 
\delta D/\pi$. 

The correlators (\ref{45}) satisfy the general relations (\ref{8}) 
and (\ref{4}), and therefore the energy sensitivity of the quantum 
point contact reaches the fundamental quantum limit (\ref{39}).  
It should be noted, however, that this conclusion is strictly valid  
only in the large-voltage limit $eV\gg T$. At finite temperature 
$T$, scattering of the point-contact electrons within the same 
direction of propagation (described by the terms $U_{11}$ and 
$U_{22}$ of the perturbation $U$) creates additional contribution 
to the backaction noise and degrades the energy sensitivity. 
The magnitude of this effect depends on the magnitude 
of the ``forward'' scattering matrix elements $U_{11}$ and 
$U_{22}$ relative to the backscattering matrix element $U_{12}$, 
increasing with intensity of the forward scattering but  
decreasing with $T/eV$ ratio.

\section{Flux and charge MQC oscillations }   

\vspace{1ex}

This Section provides specific examples of measurements of 
the macroscopic quantum coherent oscillations of magnetic flux and 
electric charge. It is shown that the typical detectors for the flux 
and charge measurements, dc SQUID and Cooper-pair electrometer, 
satisfy the general equations of Sections 2 and 3, and  
should be capable of reaching the fundamental limit of the 
signal-to-noise ratio for the continuous weak measurement of the MQC 
oscillations.

\subsection{Flux oscillations measured with a dc SQUID} 

Typical set-up of a measurement of the MQC oscillations of flux 
with a dc SQUID consists of a two-state flux system (rf SQUID with 
half of a magnetic flux quantum $\Phi_0=\pi\hbar/e$ induced in it 
by an external magnetic field) coupled inductively to a dc SQUID 
biased with an external current $I_0$ and shunted by a resistor $R$ 
(Fig.\ 3). When the inductance of dc SQUID loop is small, the 
difference between the two Josephson phases $\varphi_{1,2}$ across 
the two junctions of the SQUID is directly linked to the flux $\Phi$ 
induced in the dc SQUID by the flux oscillations: 
\[ \varphi_{1}- \varphi_{2} =2\pi \Phi/\Phi_0 \equiv \Theta\, . \]  
In this case, the SQUID is equivalent to a single Josephson 
junction, with the supercurrent in this junction modulated by the 
flux $\Phi$. The total amplitude of Cooper-pair tunneling in the 
SQUID is equal to a sum  of the two individual amplitudes of 
tunneling in the two SQUID junction, and can be written as
\begin{equation} 
E_J/2=I^{(+)}(\Theta)/4e\, ,\;\;\;\; I^{(+)}(\Theta) = 
I_1e^{i\Theta/2} + I_2e^{-i\Theta/2} \, , 
\label{51} \end{equation}
where $I_{1,2}$ are the critical currents of the two junctions. 
Coherent sum of the two tunneling amplitudes in (\ref{51}) 
leads to modulation of the total supercurrent of the SQUID. 

\begin{figure}[htb]
\setlength{\unitlength}{1.0in}
\begin{picture}(3.,1.76) 
\put(.25,.1){\epsfxsize=2.7in\epsfysize=1.6in\epsfbox{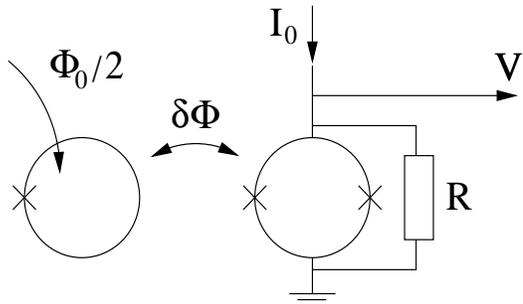}}
\end{picture}

\caption{ Schematic diagram of a continuous measurement of the 
MQC oscillations of magnetic flux 
with a dc SQUID. The two-state flux system is implemented as 
an rf SQUID with externally induced flux $\Phi_0/2$ and is 
coupled to the dc SQUID biased with a current $I_0$. The 
coupling strength is characterized by the variation $\delta 
\Phi$ of the flux through the dc SQUID due to oscillations. 
The voltage $V$ across the dc SQUID is the measurement output.  }

\end{figure}

We consider the simplest 
regime of the dc SQUID dynamics when the bias current $I_0$ 
and associated average voltage $V_0=RI_0$ across the dc SQUID are 
sufficiently large, and the dc component of the Josephson current 
through it is small in comparison to $I_0$. In this regime, the 
Cooper-pair tunneling through the SQUID is adequately 
described by perturbation theory in the tunneling amplitude 
(\ref{51}) and can be qualitatively interpreted as incoherent 
jumps of individual Cooper pairs. The resistor $R$ provides the 
dissipation mechanism that transforms reversible dissipationless 
Cooper-pair oscillations between the two electrodes of the SQUID 
into incoherent tunneling. 

Using the known results for the incoherent Cooper-pair tunneling 
\cite{b42}, we can calculate the rate of this tunneling and find 
all the detector characteristics of the dc SQUID. The SQUID 
is coupled to the oscillations by the operator of the 
current $I_-$ circulating in the SQUID loop multiplied by the 
change $\delta \Phi$ of the flux in the loop induced by the flux 
oscillations. Changes in the flux through the dc SQUID change 
the total current $I_+$ through both SQUID junctions and create 
deviations $V$ of the voltage across the SQUID from $V_0$, 
$V=-RI_+$, that serve as the measurement output. 
As follows from the general discussion in Sec.\ 
2, SQUID parameters important for measurement are the coefficient 
$\lambda$ of the transformation of the oscillating flux into the 
voltage $V$, the spectral density $S_I$ of the circulating current 
$I_-$ that is responsible for backaction dephasing by the SQUID, 
the spectral density of the output voltage $S_V$, and the 
correlator $S_{VI}$ between $V$ and $I_-$. These parameters can 
be found quantitatively starting from the tunneling part $H_T$ of 
the SQUID Hamiltonian that can be written as 
\begin{equation} 
H_T= -\frac{E_J}{2} e^{i(2eV_0t+\varphi(t))} +h.c.
\, ,  
\label{52} \end{equation}
with $\varphi(t)$ being the random Josephson phase across the dc 
SQUID accumulated due to equilibrium voltage fluctuations produced 
by the resistor $R$. It is characterized by the correlator 
\begin{equation}  
\langle \varphi (t) \varphi \rangle = \rho \int \frac{d 
\omega }{\omega} g(\omega) \frac{ e^{i\omega t } 
}{1-e^{- \omega /T} } \, , 
\label{55} \end{equation} 
where $\rho=R/R_Q$ is the resistance $R$ in units of the quantum 
resistance $R_Q=\pi \hbar/4e^2$, the average $\langle \ldots 
\rangle$ is taken over equilibrium density matrix of the resistor 
$R$, and $g(\omega)$ describes the cut-off of the dissipation 
provided by $R$ at some large frequency $\omega_c$ associated with 
either finite inductance of the SQUID or finite capacitance of its 
junctions, while $g(\omega)=1$ at $\omega \ll \omega_c$. 

The operators of the two currents $I_\pm$ that determine the 
SQUID parameters are: 
\begin{equation} 
I_\pm= \frac{-i}{2}[I^{(\pm )} (\Theta) e^{i(2eV_0t+\varphi 
(t))} - h.c.]  \, , 
\label{53} \end{equation}
\[  I^{(-)} \equiv (I_1e^{i\Theta/2}-I_2e^{-i\Theta/2})/2\, .\] 
In the regime of the incoherent Cooper-pair tunneling the average 
dc current $\langle I_+ \rangle$ can be found treating the 
tunneling $H_T$ as perturbation:
\begin{equation}   
\langle I_+ \rangle = -i\int_0^{\infty} dt \langle 
[I_+,H_T(t)] \rangle  = \pi |I^{(+)}(\Theta)|^2 \tau/e \, , 
\label{54} \end{equation}
where 
\begin{equation}
\tau\equiv \frac{1}{4\pi} \mbox{Re} \int_0^{\infty} dt e^{i2eV_0t} 
\langle [e^{i\varphi(t)},e^{-i\varphi} ] \rangle \, .   
\label{56} \end{equation} 
For example, for vanishing temperature $T$, and small bias 
voltages, $2eV_0 \ll \omega_c$, the time $\tau$ defined in 
(\ref{56}) can be found from eqs.\ (\ref{56}) and (\ref{55}) to 
be (see, e.g., \cite{b43}):  
\begin{equation} 
\tau = (1/4 \omega_c\Gamma (\rho)) (2eV_0/\omega_c )^{ 
\rho-1} \, . 
\label{56a} \end{equation}  
When the resistance $R$ is small, $R\ll R_Q$, $\tau$ becomes 
independent of $\omega_c$, $\tau=eR/2\pi V$. In this case, 
all the SQUID characteristics, including the average current 
$\langle I_+ \rangle$ (\ref{54}) can be obtained by direct time 
averaging of the classical Josephson oscillations in the SQUID.    

The noise spectral densities of the two currents, $I_\pm$, are 
obtained by directly taking the average over equilibrium density 
matrix of the resistor $R$. They vary with frequency on the scale 
of the Josephson frequency $2eV_0$, and are constant at $\omega 
\ll 2eV_0$. In this frequency range,  
\[ S_I =\frac{1}{2\pi} \int d t  \langle I_-(t)I_- \rangle = 
\frac{1}{8\pi}|I^{(-)}(\Theta)|^2 \times \] 
\begin{equation} 
\int d t e^{i2eV_0t} \langle 
[e^{i\varphi(t)},e^{-i\varphi} ]_+ \rangle \, .
\label{57} \end{equation}
Fluctuation-dissipation theorem relates the anticommutator 
$[\ldots ]_+$ in this equation to the commutator in eq.\ (\ref{56}) 
and gives:
\begin{equation}
S_I = |I^{(-)}(\Theta)|^2 \tau' \, , \;\;\; S_V = R^2 
|I^{(+)}(\Theta)|^2 \tau'\, ,   
\label{58} \end{equation} 
where $\tau' \equiv \tau \coth (eV_0/T)$. 
The correlation function $S_{VI}$ is found similarly: 
\begin{equation} 
S_{VI} = R [I^{(+)}(\Theta)]^*I^{(-)} (\Theta) \tau'\, .
\label{59} \end{equation}
Comparison of the spectral density $S_V$ (\ref{58}) and the 
average current (\ref{54}) shows that $S_{I_+}=S_V/R^2= (e 
\langle I_+ \rangle /\pi)\coth (eV_0/T)$, i.e. the noise of the 
current $I_+$ can indeed be interpreted as resulting from 
uncorrelated transitions of individual Cooper pairs. In 
particular, at $T\ll eV_0$, the noise is the shot noise of 
Cooper pairs.    

Finally, eq.\ (\ref{54}) gives the response coefficient of the  
SQUID  
\begin{equation} 
\lambda \equiv \partial V/\partial \Phi = 2\pi R 
(\partial |I^{(+)}(\Theta)|^2/ \partial \Theta) \tau 
\, . 
\label{60} \end{equation}
(Note that as in eq.\ (\ref{58}) for the backaction noise $S_I$ 
and also in eq.\ (\ref{59}) for the correlator $S_{VI}$, the 
factor $\delta \Phi$ is omitted from the definition of $\lambda$.)  
 
For temperatures negligible on the scale of $eV_0$, $\tau'$ in 
the noise spectral densities is equal to $\tau$ and eqs.\ 
(\ref{58}) through (\ref{60}) show that the spectral densities 
$S_V$, $S_I$, and $S_{VI}$ satisfy the general relation (\ref{8}),   
and since $\partial |I^{(+)}(\Theta)|^2/ \partial \Theta = - 2 
\mbox{Im} \{ [I^{(+)}(\Theta)]^* I^{(-)} (\Theta)\}$, the 
correlator $S_{VI}$ is also related to the response coefficient 
$\lambda$ by the expression identical to eq.\ (\ref{4}). Moreover, 
since 
\[ [I^{(+)}(\Theta)]^*I^{(-)} (\Theta)=\frac{1}{2}(I_1^2-I_2^2)+ 
i I_1I_2 \sin \Theta \, , \]
we see that the real part of the correlator $S_{VI}$, which 
increases backaction dephasing produced by the SQUID, is indeed 
associated with the SQUID asymmetry. When $I_1=I_2$, the real 
part vanishes and the SQUID as detector reaches quantum-limited 
optimum for measurement of the quantum coherent oscillations. 
For such a symmetric SQUID the signal-to-noise ratio of the 
oscillation measurement is given by eq.\ (\ref{40}) (with the 
intensity of background output noise given by $S_V$). If the 
temperature $T$ of the symmetric SQUID is negligible, eqs.\ 
(\ref{58}) through (\ref{60}), and (\ref{38}) show that in 
this regime the SQUID is the quantum-limited detector with the 
energy sensitivity $\epsilon=\hbar/2$, and the signal-to-noise 
ratio for measurement of the quantum flux oscillations is 
$S_{max}/S_V=4$. When $T$ becomes non-vanishing, both the output 
and backaction noise increase, $\tau'>\tau$, and eq.\ (\ref{40}) 
describes the gradual suppression of the signal-to-noise ratio 
with increasing temperature.

\subsection{Charge oscillations measured with a Cooper-pair 
electrometer} 

Coherent oscillations of charge take place in Josephson junctions 
which are sufficiently small for the charging energy $E_C$ of an 
individual Cooper pair, $E_C=(2e)^2/2C$ to be larger than  
temperature $T$ and Josephson coupling energy $E_J$. The 
supercurrent flow through the junction is ``discretized'' in this 
regime into the transfer of individual Cooper pairs by strong 
Coulomb repulsion. Quantitatively, if the charging energy is 
smaller that the superconducting energy gap $\Delta$ 
so that the dissipative quasiparticle tunneling is suppressed, 
the junction dynamics is governed by the simple Hamiltonian: 
\begin{equation}  
H=E_C (n-q)^2-\frac{E_J}{2}(|n\rangle\langle n+1|+|n+1\rangle 
\langle n| ) \, , 
\label{70} \end{equation} 
where $n$ is the number of Coper-pairs charging the junction, and,  
here and below, $q$ is the charge (in units of 2$e$) injected into 
the junction from external circuit. Eigenstates of the Hamiltonian 
(\ref{70}) form energy bands as functions of the injected charge 
$q$, which can be varied continuously. Variations of the injected 
charge $q$ within these bands leads to the possibility of 
controlling the tunneling of individual Cooper pairs \cite{b44}. 

The best way of injecting the charge $q$ in a junction is 
provided by the ``Cooper-pair box'' system \cite{b45,b46} in which 
the junction is attached to external bias voltage $V_g$ through 
a capacitor. If $q$ is fixed at half of a Cooper-pair charge, 
$q\simeq 1/2$, and the tunneling amplitude $E_J/2$ is much less 
than $E_C$, the two states of the Hamiltonian (\ref{70}): $n=0$ 
and $n=1$ are nearly-degenerate and separated by the large energy 
gaps from all other states. In this regime, the junction dynamics 
is equivalent to that of a regular quantum two-state system with the 
two basis state that correspond to a Coooper pair being on the 
left or on the right electrode of the junction. Coherent 
superposition of charge states in such a two-state system is 
observed indirectly by measuring either the width of the transition 
region between the two charge states \cite{b47} or the energy gap 
between the eigenstates \cite{b48} as functions of the induced 
charge $q$. Quantum coherent oscillations between the two charge 
states were also observed directly in the time-dependent 
measurement \cite{b10}. 

\begin{figure}[htb]
\setlength{\unitlength}{1.0in}
\begin{picture}(3.3,1.15) 
\put(-.15,.1){\epsfxsize=3.5in\epsfysize=1.1in\epsfbox{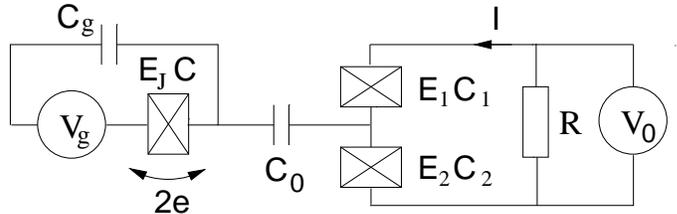}}
\end{picture}

\caption{Schematic diagram of a continuous measurement of the 
quantum oscillations of a Cooper-pair in a small Josephson 
junction biased with charge $V_gC_g\simeq e$ through a small 
capacitance $C_g$. Capacitance $C_0$ couples the oscillating Cooper 
pair weakly to an electrometer composed of two junctions with 
Josephson coupling energies $E_{1,2}$ and capacitances $C_{1,2}$. 
The electrometer is biased with a voltage $V_0$, while the 
tunneling current $I$ through it is the measurement output.}

\end{figure} 

The experiment \cite{b10} was effectively based on the strong 
measurement of charge oscillations, when each measurement 
suppresses the oscillations, and they are observed as 
oscillations of probability in an ensemble of measurements.   
Continuous weak measurement of the quantum-coherent charge 
oscillations similar to the measurement of the flux 
oscillations with a dc SQUID discussed above would provide a less 
intrusive way of studying these oscillations. One of the detectors 
appropriate for such a measurement is a Cooper-pair electrometer 
\cite{b31,b49}: two small Josephson junctions with Josephson 
coupling energies $E_{1,2}$ and capacitances $C_{1,2}$ connected 
in series and shunted with a resistor $R$ (Fig.\ 4). 
As in the case of dc SQUID, we consider dynamics of the Cooper 
pair transfer through the electrometer in the regime 
of incoherent tunneling. The main difference with the SQUID case 
is that now the amplitude of Cooper pair tunneling is modulated 
through the modulation of energy of the intermediate state in 
the process of the two-step transfer of Cooper pairs in the two 
junctions. At small bias voltages $V_0\ll E_C/e$, where from now 
on $E_C=2e^2/(C_1+C_2)$ is the charging energy of the central 
electrode of the electrometer, the intermediate state is virtual,  
and the average current $\langle I\rangle$ through the electrometer 
is determined by the same eq.\ (\ref{54}) with $I^{(+)}(\Theta)$ 
replaced with the amplitude $I(q)$ of Cooper-pair transfer through 
both junctions (defined below). The charge $q$ injected into the 
central electrode controls the energy of the intermediate states 
in the process of the Cooper-pair transfer and modulates the 
tunneling amplitude. At small Josephson coupling energies 
$E_{1,2}\ll E_C$, and away from the resonance points $q=\pm 1/2$, 
Cooper-pair tunneling can be treated as perturbation. Then the 
instantaneous value of the current $I$ for a fixed value of the
Josephson phase $\varphi$ across the electrometer (see, e.g., 
description of the two-junction system in \cite{b50}) is: 
\begin{equation} 
I= I(q)\sin (2eV_0t+\varphi(t))\, , 
\label{71} \end{equation}
\[ I(q)\equiv \frac{eE_1 
E_2 }{ E_C} (\frac{1}{1-2q}+\frac{1}{1+2q})\, . \]
The two terms in the second equation in (\ref{71}) correspond to 
the two intermediate states with different charges $n=\pm 1$ on the 
central electrode of the electrometer in the Cooper-pair transfer 
process. 

Averaging eq.\ (\ref{71}) over the equilibrium quantum fluctuations 
of $\varphi$ we get expression for the average value of the current 
$I$ that is equivalent to eq.\ (\ref{54}). Since the tunneling 
current $I$ through the electrometer is the measurement output, this 
expression determines the response coefficient $\lambda$ of the 
electrometer: 
\begin{equation}  
\lambda \equiv (\partial \langle I \rangle/ \partial q)= 
\pi (\partial [I(q)] ^2/ \partial q) \tau/e \, , 
\label{72} \end{equation}
where $\tau$ is given by eqs.\ (\ref{56}) and (\ref{56a}). 
Similarly, the output noise of the current $I$ can be obtained as 
\begin{equation}
S_I = [I(q)]^2 \tau' \, . 
\label{73} \end{equation} 

The backaction noise of the Cooper-pair electrometer is created by  
fluctuations of the charge on its central electrode in the process 
of the Cooper-pair tunneling. This fluctuations lead to fluctuations 
of electric potential of this electrode. In the same regime as for 
eq.\ (\ref{71}), the magnitude of this fluctuations is determined 
by the magnitude of the instantaneous value of the potential at a 
fixed Josephson phase difference $\varphi$ across the electrometer: 
\begin{equation} 
V= U(q)\cos (2eV_0t+\varphi(t))\, , 
\label{74} \end{equation}
where
\[ U(q)\equiv \frac{E_1E_2 
}{2e E_C} (\frac{1}{(1-2q)^2}-\frac{1}{(1+2q)^2})\, . \]
As before, averaging over the fluctuations of $\varphi(t)$ we get 
the spectral density of the backaction noise and its correlation  
with the output noise: 
\begin{equation} 
S_V = [U(q)]^2 \tau'\, , \;\;\; S_{VI} = -i U(q) I(q) \tau'\, . 
\label{75} \end{equation}
Equations (\ref{72}), (\ref{73}), and (\ref{75}) show that 
at vanishing temperature, when $\tau'= 
\tau$, the noise characteristics of the Cooper-pair electrometer 
satisfy the general relations (\ref{8}) and (\ref{4}) of a  
quantum-limited detector. Moreover, the electrometer is 
``symmetric'' detector in a sense that the input-output correlator 
$S_{VI}$ is purely imaginary. This means that both its energy 
sensitivity $\epsilon$ and the signal-to-noise ratio (\ref{40}) for 
measurement of the two-state system reach fundamental limits. 

In summary, the examples of specific detectors considered in this 
Section show explicitly that many standard detectors  
should be capable of reaching the fundamental limits of sensitivity 
for measurements of electric charge and magnetic flux. In this 
regime, they are characterized by the fundamental signal-to-noise 
ration of 4 for continuous weak measurement of the macroscopic 
quantum coherent oscillations. The limitation on the signal-to-noise 
ratio of such a measurement has the same origin as the quantum 
limitation on 
the operation of an ideal linear phase-insensitive amplifier that 
adds a minimum of half-a-quantum of noise to the amplified signal.

The author would like to acknowledge discussions with M.H. Devoret, 
J.R. Friedman, A.N. Korotkov, K.K. Likharev, J.E. Lukens, Yu.V. 
Nazarov, R.J. Schoelkopf, G. Sch\"{o}n, and A.B. Zorin. This work 
was supported in part by AFOSR. 

\vspace*{2ex} 

{\bf Appendix} 

\vspace*{1ex} 

In this Appendix, we derive eq.\ (\ref{43}) that relates the 
matrix elements of the perturbation of the scattering potential 
to the transmission properties of a point contact. Relation (\ref{43}) 
can be established considering the stationary states of an electron 
confined to move on the interval $x\in [-L/2,L/2]$ with the main 
scattering potential located at the center of the interval, $x\simeq 0$. 
The scattering matrix $S$ for the symmetric scattering potential can be 
written as 
\[ S=e^{i\Theta} \left( \begin{array}{cc} i\sin \nu \, , & \cos \nu \\
\cos \nu \, , & i\sin \nu \end{array} \right) \, , \] 
where $\Theta$ is the phase of the transmission amplitude, and $\nu$ 
parametrizes transmission probability $D$: 
\begin{equation}
D = \cos^2 \nu\, . 
\label{a1} \end{equation}
Diagonalizing the scattering matrix $S$, we find the two phase shifts 
$\varphi_{1,2}$ associated with it: $\varphi_1= \Theta +\nu$, 
$\varphi_2= \Theta +\pi -\nu$. The variations $\delta \varphi_j$ of 
the two phase shifts due to perturbation of the scattering potential 
lead to changes $\delta \varepsilon_j$ in energies of the two 
stationary states, 
\begin{equation} 
\delta \varepsilon_j = v_F \delta \varphi_j/L 
\label{a2} \end{equation}

For symmetric main scattering potential, the two stationary states 
$\chi_j(x)$ are given by the even and odd combinations of the 
scattering states $\psi_j(x)$. In the basis of the states $\chi_j$, 
the perturbation matrix $U$ introduced in eq.\ (\ref{41}) is: 
\begin{equation}
U=\left( \begin{array}{cc} (U_{11}+U_{22})/2+\mbox{Re} U_{12} 
\, , &  (U_{11}-U_{22})/2-i\mbox{Im} U_{12}\\
(U_{11}-U_{22})/2+i\mbox{Im} U_{12}\, , & 
(U_{11}+U_{22})/2-\mbox{Re} U_{12} \end{array} \right) \, . 
\label{a3} \end{equation}
The diagonal elements of this matrix give the first-order 
corrections to the energies of the stationary states. Comparing 
expressions for the energy corrections given by eq.\ (\ref{a3}) 
to expressions for the phase shifts combined with eq.\ (\ref{a2}), 
we see that  
\[ \mbox{Re} U_{12} = v_F \delta \nu /L \, . \] 
This equation, together with the relation (\ref{a1}) between 
$\nu$ and transmission probability $D$, gives eq.\ (\ref{43}) 
of the main text. Since the matrix (\ref{a3}) should be symmetric 
for the perturbation of the scattering potential that is symmetric 
with respect to the main part of the potential, we also see that 
$\mbox{Im} U_{12} =0$ in the symmetric case. This means that the 
nonvanishing imaginary part of $U_{12}$ can be viewed as a 
measure of the asymmetry of coupling between the point contact 
and a source of the perturbation.

\end{document}